# Cultural Differences in E-Learning: Exploring New Dimensions

NAZIA HAMEED [1], MAQBOOL UDDIN SHAIKH[2], FOZIA HAMEED[3], AZRA SHAMIM[4]

[1]Computer Science Department, COMSATS Institute of Information Technology, Islamabad, Pakistan
nazia_hameed@comsats.edu.pk

[2]Computer Science Department, PRESTON University, Islamabad, Pakistan
maqboolshaikh@preston.edu.pk

[3]Computer Science Department, KING KHALID University, Saudi Arabia
fozi_aug12@yahoo.com

[4]Faculty of Computer Science and Information Technology, University of Malaya, Malaysia
azra.majeed864@yahoo.com

*Abstract:* - Rapid development of Internet and information technologies has gifted us with a new and diverse mode of learning known as e-learning. In the current era, e-learning has made rapid, influential, universal, interactive, vibrant, and economic development. Now e-learning has become a global mode of education. E-learning means the use of internet, computer and communications technologies to acquire education. Learners with diverse social, cultural, economic, linguistic, and religious backgrounds from all over the world are taking benefits from e-learning. In e-learning, culture of target learners plays a vital role and need to be explored for better results. Diversity of culture and learning styles should keep under considerations while designing e-learning environment. It may help to attain the required results. In this research work, authors proposed and designed a novel architecture for e-learning system incorporating cultural diversity of learners. The focus is to concentrate on cultural factors from e-learning system. Furthermore; a prototype of the proposed system is implemented for the validation of proposed architecture.

## 1 Introduction

The most promising outcome of the Internet and information technology is e-learning. E-learning has made rapid development and has become a global mode of education. It is an umbrella term that describes learning done at a computer, online or offline. E-learning is the process of delivering knowledge to the learner with the help of computers, Internet, Intranet, and the Web. In e-learning process there are many issues need to be further investigated. One of the issues that need to be explored is cultural differences [1][2]. It is an umbrella term that includes living life style, education system, place, gender, race, history, nationality, language, sexual orientation, religious beliefs, ethnicity and aesthetics.

Pakistan is a developing country occupying a vital geographical position and has culturally diverse regions. In Pakistan, there are many educational institutes having different educational environment. Student enrolled in e-learning system belongs to different regions have different culture values. Their education system, medium of instruction, course contents etc. are different. Students of backward areas don't have good exposure of information and communication technology (ICT). Most of existing e-learning systems do not consider these differences due to which learners face many problems during their study. Cultural differences are major problem which effect e-learners [1],[3],[4],[5],[6]. Current e-learning system does not consider cultural diversity of learners. In this paper, authors designed and proposed an e- learning system that caters the cultural difference issue of learners. Furthermore, for verification authors have developed a prototype of proposed system.

The rest of the paper is organized as follows. Section 2 illustrates the literature review; Section 3 describes the proposed architecture. Section 4 presents results and discussions.

## 2 Literature Review

### 2.1 e-learning
According to Badrul H. Khan, "E-learning can be viewed as an innovative approach for delivering well-designed, learner-centered, interactive, and facilitated learning environment to anyone, anyplace, anytime by utilizing the attributes and resources of various digital technologies along with other forms of learning materials suited for open, flexible, and distributed learning environment" [6]. E-learning is also defined as "e-learning is the use of technology to enable people to learn anytime and anywhere. E-learning can include training, the delivery of just-in-time information and guidance from experts" [7].

## 2.2 Mode of e-learning

There are two modes of e- learning, i.e. synchronous and asynchronous. In synchronous mode, classes take place in a classroom in real-time e.g. communication between instructors and learners through teleconferencing. In asynchronous mode, learners can access educational material in their convenient time anywhere. In synchronous mode, learning takes place in real time but in asynchronous, it is not take place in real time. A brief description of these modes is presented in[11], [12].

## 2.3 Modalities of e-learning Activities

According to Som Naidu, there are four types of e-learning activities i.e. individualized self-paced e-learning online, individualized self-paced e-learning offline, group-based e-learning synchronously; and group-based e-learning asynchronously [13]. In individualized self-paced e-learning online; learners use the internet for accessing learning resources. In individualized self-paced e-learning offline, learners access learning resources without the Internet. In group-based e-learning synchronously, groups of learners are working together in real time via the Internet. In group-based e-learning asynchronously, groups of learners are working over the Internet but not in real time [9],[13].

## 2.4 E- learning Models

Several e-learning models are proposed by different researchers [6],[8],[9],[13]. Some models emphasis on the interactions between the teacher, student and the content [5],[8] while others focus on embedding tacit knowledge in pedagogical model to enhance e-learning [13].

Khan proposed P3 model for e-learning in [6]. The content development phase involves planning, design, development and evaluation of e-learning content and resources. Implementation of online course offerings, monitoring and updating of e-learning environment are included in delivery and maintenance phase.

Terry et. al. in [8] proposed a model of e-learning in which different types of interactions are illustrated. Two major human characters of e-learning i.e. teachers and student are represented in the model. Learners can directly interact with the content as the e-learning material can be easily access able from the internet. Rizwana et. al. in [14] proposed a structure for embedding tacit knowledge in pedagogical model. Her proposed pedagogical framework includes following three phases; content organization, quality assessment and content delivery phase. In the contents organization phase, the contents used for learning are collected and organized based on the requirements put forward by the industries. Quality assessment phase analyzes and assesses the contents developed and gathered based on theoretical and experimental methods. In contents delivery phase, the learner is assessed to find out their aptitude and level.

## 2.5 Culture Difference in e-learning

Culture can be defined as "the beliefs, value systems, norms, mores, myths, and structural elements of a given organization, tribe, or society" [15].Culture affects the learner's behaviors of online education and these effects must be taken into consideration to make e-learning more effective and practical.

## 2.6 Effects of Culture on the Design and Development of e-learning System

E-learning systems are build while considering needs and the customs of learners. Without keeping these considerations the system cannot be effective and thus unable to attain the required results. To analyze the differences, the communication disciplines involved are computer-mediated communication, cross-cultural communication cultural, context interaction mode and communication form [16].

With the advent of new technologies in distributed learning, e-learning courses can be accessed throughout the global world. The diversity of culture and learning styles should kept under considerations while designing e-learning system. One of the difficult issues in e-learning is cross-cultural communication [17].

Eight educational value differentials are highlighted by Bentley et. al. in [18] to understand the cultural issues. These issues are language differential, educational culture difference, technical infrastructure differential, local versus global differential, learning style differential, reasoning pattern differential, high and low-context differential and social context differential. The authors not only incorporate the language and technology but also learning styles and local and global context.

# 3 Proposed Architecture

The proposed architecture is discussed below and shown in figure 1
.

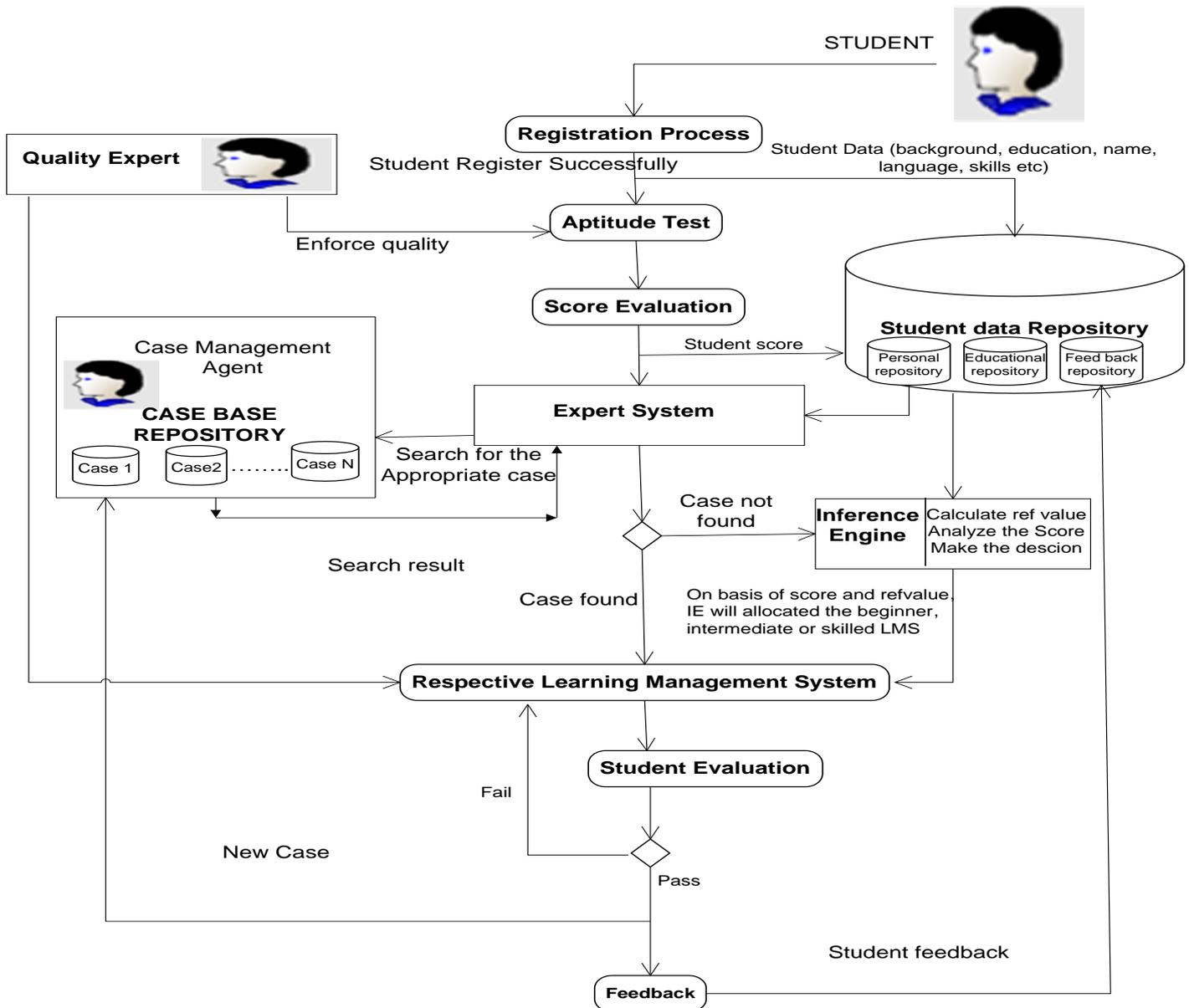

Figure 1: Proposed Architecture

### 3.1 Registration Process
In registration process, student registers a course through e-learning system. In first step student is asked for signup. At signup student enters two types of data; personal and cultural. Student data is stored in student data repository which is divided into three sub repositories i.e. personal repository, cultural repository and feedback repository. Student personal information is stored in personal repository. Whereas student culture information like school type i.e. government or private, medium of instruction of primary secondary, course contents of previous study, area from which student belongs, computer skills, school environment, economic background etc are stored in cultural repository.

Students feed backs are stored in this repository for future decision making.

### 3.2 Aptitude Test
Aptitude test is used to judge the student knowledge level. After signing in student take the aptitude test. Quality is a major concern in designing of the aptitude test. In our system quality manager assure that the questions designed for aptitude test are according to educational standards. Aptitude test consists of four portions; English, Mathematical reasoning, Computer and Intelligence Quotient. Each portion consists of ten questions and each question carry one point. Aptitude test starts with English portion. When English portion is successfully completed, students solve mathematical reasoning, computer and intelligence quotient portions respectively.

### 3.3 Aptitude Test Evaluation

When aptitude test is successfully submitted, the score of English ($S_E$), Mathematical Reasoning ($S_{MR}$), Computer ($S_C$) and Intelligence Quotient ($S_{IQ}$) is evaluated individually.

Total = $S_E + S_{MR} + S_C + S_{IQ}$  (1)

Total score of aptitude test is calculated using equation 1. After evaluating total, percentage is calculated. Total score and percentage of a student's aptitude test is then stored into in the student repository.

### 3.4 Inference Engine

Inference engine check the educational background of the student and calculate the reference value ($refvalue_{student}$) of the student. $refvalue_{student}$ of student depend on the cultural and educational information. $refvalue_{student}$ is based on the following factors:

- Learning language (Medium of Instruction)

Medium of instruction is a key factor that should be considered in e-learning system. Students who have English language as medium of instruction perform better then the students who have Urdu language as the medium of instruction.

- Computer Knowledge

An important factor in calculation of $refvalue_{student}$ is student's knowledge about computer. Marzano et. al. in [19] stressed the importance of having background knowledge and concluded that insufficient background knowledge causes lower achievement in learners.

- Course Contents (Local or International)

Another important factor is course contents of the student's previous study. Contents of the course and the language affect a lot on the learner knowledge [20]

- Percentage Marks of Aptitude Test

Other factor considered for $refvalue_{student}$ is student's percentage marks of aptitude test. Levels are assigned to the students and level depend on $refvalue_{student}$ and .

There are three levels in which students are categorized listed below: Beginner, Intermediate and Skilled

The assignment of levels is summarized in the pseudo code below:

IF (((RA=7)&(%>=60))||((RA=6) & (%>=70))|| (( RA=5)&(%>=80))||(( RA=4) & (%>=85))||((RA=3) &(%>=90)))
Then Level = Skilled
Else IF ((( RA=7) &((%<60) &(%>=50))||(( RA=6) & ((%<70) & (%>=60))) ||(( RA=5) & ((%<75) &(%>=60)) ||(( RA=4) & ((%<85)& (%>=70))) ||(( RA=3)& ((%<95) & (%>=80)))
Then Level = Intermediate
Else IF ((( RA=7)&((%<50)&(%>=40))|| (( RA=6) & ((%<50) &(%>=40))) ||(( RA=5) & ((%<60) & (%>=40)) || (( RA=4) &((%<70) & (%>=40))) ||(( RA=3)&((%<80) & (%>=40)))
Then Level = Beginners
Else IF ( %<40)
Then not eligible for the degree
End IF
(RA= Average Reference Value)

### 3.6 Learning Management System

After the levels have been assigned to the students, student is allocated to the respective learning management system (LMS) as shown in Figure 2. "LMSs facilitate learning by providing a centralized location were learning material reside and through integrated tools that enable various teaching and learning activities such as communication and collaboration with peers and lecturers, self-assessment and progress tracking" [21].

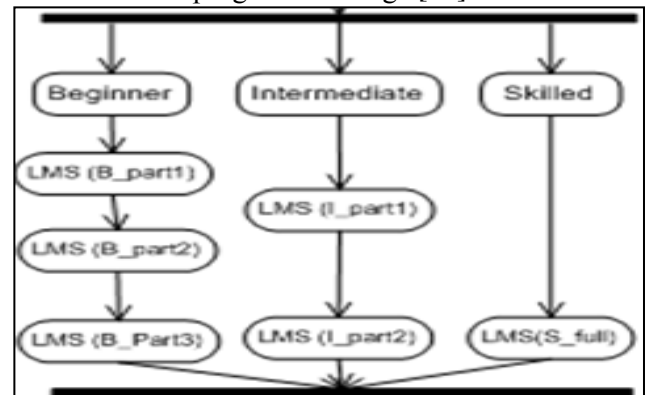

Figure 2: LMS for Beginners, Intermediate and Skilled

### 3.7 Student Evaluation

Students are evaluated after they successfully complete the course. During their course they are evaluated frequently with assignments and quizzes. If the students fail in course evaluation they must take the course again.

### 3.8 Case base Reasoning

Case base reasoning is an additional feature in the proposed architecture. Every student who successfully passes the course is stored as new case in case base repository with all the relevant data.

### 3.9 Feedback

Feedback from students should be considered important in to e-learning systems. Feedback form is evaluated from students who successfully pass the course. Their suggestions are stored in the feedback repository. These suggestions are helpful for making future improvements in the system.

## 4 System Implementation

In the working prototype, there are two interfaces

(1) Admin end and (2) User end. Admin end is created to facilitate the administrator of the system. Using admin panel; administrator can manage aptitude test. Administrator can easily add, modify and delete questions in the test. Admin panel is shown in figure 3

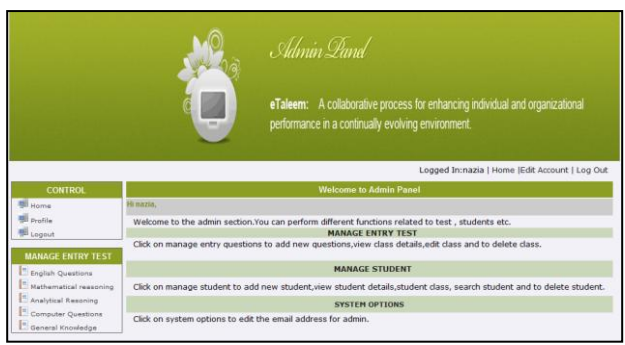

Figure 3: Admin Panel

User end is used by students for registration. First users have to register themselves before enrolling any course. Reference value is calculated using educational and cultural information provided by the student. Then students take the aptitude test by clicking on the start button. On the completion of aptitude test, student level is calculated based upon the reference value and the marks obtain by the student After assigning the level, student will directed to respective LMS.

## 5. Results

The prototype is evaluated from different users. Authors evaluated the prototype from students of Islamabad Model College for girls, (IMCG) F6-2, Islamabad and COMSATS University, of Science and Technology, Islamabad. Among participants, 54% students are male and 46% are female as shown in Figure 3.

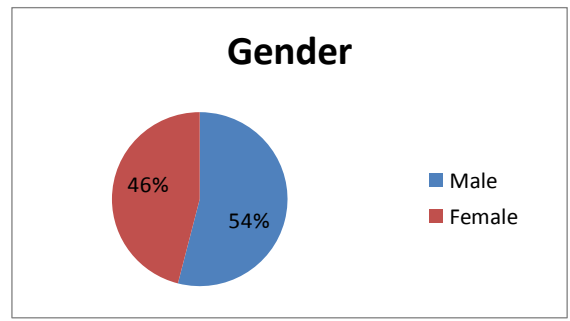

Figure 4: Percentage of Male and Female Students

Students categorized into different levels as shown in Figure 5 prove that the educational factors affect e-learning.

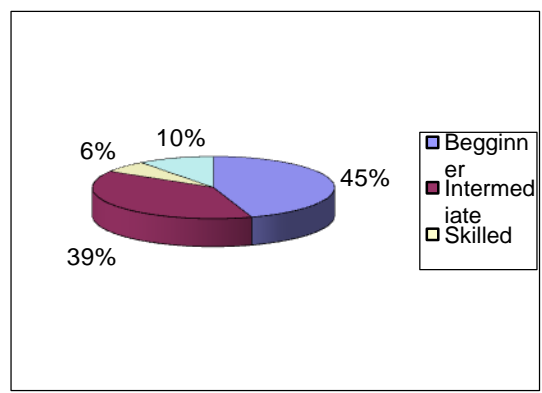

Figure 5: Effects of Education Factors on e-learning

Therefore the student's cultural diversity should be considered while offering e-learning programs. Majority of students having Urdu language as their medium of instruction falls in the beginner level, Hence we can conclude from these results that the medium of instruction affect a lot on student learning as shown in Figure 6.

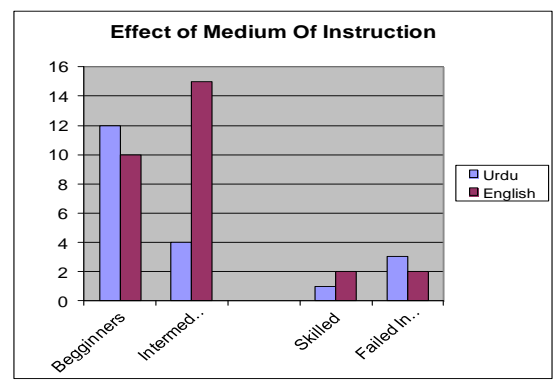

Figure 6: Effect of Medium of Instruction

Results gathered from the proposed architecture shows that the students who have studied international course contents have better score then those who studied the local course contents, which concludes that course contents also affects the learners.

Students who don't have exposure to computer face many problems during e-learning therefore Human Computer Interaction and its principles should be considered while designing systems for students having different educational culture.